\documentclass[12pt]{article}
\usepackage{amsmath}
\usepackage{graphicx,psfrag,epsf}
\usepackage{enumerate}
\usepackage[numbers]{natbib}
\usepackage{textcomp}
\usepackage[hyphens]{url} 
\usepackage[hidelinks]{hyperref}
\usepackage{lmodern}
\usepackage[parfill]{parskip}
\usepackage[justification=centering]{caption}
\usepackage{booktabs}
\usepackage{longtable}


\usepackage{mathptmx}

\usepackage{graphics}
\usepackage{amsthm}
\usepackage{amssymb}
\usepackage{bm}

\begin{document}

\def\spacingset#1{\renewcommand{\baselinestretch}%
{#1}\small\normalsize} \spacingset{1}

\title{\bf NFL Injuries Before and After the 2011 Collective Bargaining Agreement (CBA)}

  \author{
   Zachary O. Binney, PhD\textsuperscript{1}; Kyle E. Hammond, MD\textsuperscript{2}; Mitchel Klein, PhD\textsuperscript{1};\\
   Michael Goodman, MD MPH\textsuperscript{1}; A. Cecile J.W. Janssens, PhD\textsuperscript{1}\\
   \\
   \textsuperscript{1}Department of Epidemiology, Rollins School of Public Health, Emory University;\\
   \textsuperscript{2}Department of Orthopedics, Emory University School of Medicine\\
      }

\maketitle

\begin{abstract}
\noindent\textbf{Introduction}: The National Football League’s (NFL) 2011 collective bargaining agreement (CBA) with its players placed a number of contact and quantity limitations on offseason, training camp, and regular season practices and workouts. Some coaches and others have expressed a concern that this has led to poor conditioning and a subsequent increase in injuries. Rigorous studies on the effects of the NFL’s new practice restrictions have not been performed, however.\\
\textbf{Objective}: We sought to assess whether the 2011 CBA’s practice restrictions affected the number of overall, conditioning-dependent, and/or non-conditioning-dependent injuries in the NFL or the number of games missed due to those injuries.\\
\textbf{Methods}: The study population was player-seasons from 2007-2016 for any player who had participated in at least one career regular season game. We included only regular season, non-illness, non-head, game-loss injuries. Injuries were identified using a database from the website Football Outsiders based on public NFL injury reports and the injured reserve list. The primary outcomes were overall, conditioning-dependent and non-conditioning-dependent injury counts by season.  We also investigated games missed due to these injuries as a secondary outcome. We calculated injury counts and games missed per season and compared the results before (2007-2010) and after (2011-2016) the CBA. We also used a Poisson interrupted time series model to assess whether there was an immediate change after the CBA or if a pre-CBA increase in injuries accelerated post-CBA.\\
\textbf{Results}: The number of game-loss regular season, non-head, non-illness injuries grew from 701 in 2007 to 804 in 2016 (15\% increase). The number of regular season weeks missed exhibited a similar increase. Conditioning-dependent injuries increased from 197 in 2007 to 271 in 2011 (38\% rise), but were lower and remained relatively unchanged at 220-240 injuries per season thereafter. Non-conditioning injuries decreased by 37\% in the first three years of the new CBA before returning to historic levels in 2014-2016. Poisson models for all, conditioning-dependent, and non-conditioning-dependent game-loss injury counts did not show statistically significant detrimental changes associated with the CBA.\\
\textbf{Conclusions}: We did not observe a sustained increase in conditioning- or non-conditioning-dependent injuries following the 2011 CBA. Other concurrent injury-related rule and regulation changes limit specific causal inferences about the practice restrictions, however, and further studies are warranted.\\

\end{abstract}

\noindent%
\vfill

\newpage


\section{Introduction}

Periodically the National Football League (NFL) and the NFL Players’ Association (NFLPA) negotiate a new collective bargaining agreement (CBA) that governs labor associations between the NFL and its players. The most recent CBA was negotiated after a 4.5-month offseason “lockout” period prior to the 2011 season. During the lockout players and teams were prohibited from any contact including practices and medical examinations. The new CBA placed a number of limitations on offseason, training camp, and regular season practices and workouts \cite{NFLPA11}. These new limitations reduced the physical burdens put on players outside of games: organized team activities (OTAs) each spring were reduced from 14 days to 10, the voluntary offseason workout program reduced from 14 to 9 weeks, twice-daily padded practices were eliminated during the 6-week training camp prior to each season, and regular season padded practices were restricted from no limit to 14 during the 17-week season \cite{Rosenthal11, Clark15}. These regulations were implemented to improve player safety and reduce injuries by avoiding overtraining and increasing rest \cite{NFLPA11, Rosenthal11, Clark15}. Some, however, have argued that they have had the opposite effect, leading to undertraining that worsens player conditioning and leaves them more susceptible to injuries \cite{Doherty15, Florio13, Farrar13, Chao15}. 

There is evidence for both undertraining and overtraining increasing injury risk. A substantial body of sports research in track and field, baseball, cricket, rugby, and Australian Rules Football suggests that high intensity training and high activity loads may lead to injuries, particularly soft tissue injuries \cite{Gabbett16a, Windt17, Colby14, Cross16, Gabbett04, Huxley14, Gabbett12, Fleisig11, Orchard09, Gabbett11, Lee01}. On the other hand, the extant literature suggests that too little training – particularly a low long-term “chronic” workload – is also a predictor of increased injury risk in sports such as cricket and rugby \cite{Gabbett16a, Cross16, Dennis03}. The goal is to find a balance between undertraining and overtraining that leaves athletes equipped to perform at a high level and renders them less susceptible to injuries. More recent research has demonstrated the importance of a balanced “acute:chronic workload ratio,” wherein a high long-term chronic workload is necessary to achieve fitness and maximize competitive performance but excessive short-term acute workloads can lead to fatigue overwhelming fitness and an increase in injury risk \cite{Gabbett16a}. Associations between rapid changes in workload and increased injury risk have been reported in in cricket, rugby, soccer, and Australian Rules Football \cite{Cross16, Rogalski13, Ehrmann16, Hulin14, Hulin16}. 

Studies of practice or training load and injury risk in American football, however, are relatively limited. One study from the Big 10 Conference in college football found that 1998 rules to limit scrimmages and full-contact practices in the spring did not decrease the number of spring injuries per year, though injuries in the fall did decline by one third \cite{Albright04}. Two studies of middle school and high school football players in Oklahoma City found that the extent of participation in preseason conditioning activities did not differ among players who sustained an injury and those who remained injury-free \cite{Turbeville03a, Turbeville03b}. To our knowledge no similar studies of preseason practice load and injury risk been performed for the NFL, however.

It is unknown whether the NFL’s practice regimens resulted in overtraining in the pre-CBA era, undertraining in the post-CBA era, both, or neither. One research group identified a sudden increase of Achilles tendon injuries immediately after the 2011 NFL lockout ended and training camp began, noting 12 Achilles tendon ruptures the first 29 days after the lockout versus 6 and 10 total Achilles tendon ruptures the last two full seasons, respectively \cite{Myer11}. The evidence we can take from such a short period is limited, however. 

We hypothesized that the effects of undertraining and overtraining would be most apparent in overuse injuries such as muscle strains, which we refer to as “conditioning-dependent” injuries. However, fewer practices – and especially fewer padded practices – also result in fewer chances for suffering an injury in the post-CBA era. This effect may be more evident in contact-based injuries such as fractures and trauma of internal organs, which we refer to as “non-conditioning-dependent” injuries.

In an effort to tease out the varying effects of the 2011 CBA’s practice restrictions, we sought to investigate whether the restrictions were followed by changes in the number of overall, conditioning-dependent, and non-conditioning-dependent injuries in the NFL. A secondary objective was to investigate whether the 2011 CBA’s practice restrictions were associated with changes in the number of games missed due to these injuries.

\section{Methods}

\subsection{Study Population and Data Sources}
\label{sec:study_pop}

Data on injuries comes from a database maintained by the football analytics website Football Outsiders \cite{FO17}. The data have been collected prospectively since the 2007 regular season. It is based on the official public injury reports released weekly by each NFL team, supplemented with additional details from media reports where available. Information for all injuries included player name, team, position, week, season, injury type, final practice report status, and the player’s participation in that week’s game. Player-season-level data such as age, height, weight, and the total number of games played was also provided by Football Outsiders; this was linked with the injury database to generate injury counts for each player-season.

The study population included any player who had participated in at least one regular season NFL game in their career through the 2016 season. The study population included 19,803 player-seasons and 22,331 injuries from 2007-2016. We excluded 2,643 preseason injuries (11.8\%) to ensure comparable populations in the pre- and post-CBA eras since the CBA increased preseason roster sizes. We excluded 11,399 non-game-loss injuries (51.0\%) to account for the more complete reporting of minor injuries in recent years \cite{Binney17a, David12}. We excluded 685 head injuries (4.1\%) to minimize the impact of improved reporting and diagnosis of concussions on our results \cite{Currie17, Kerr17}. Finally, we excluded 179 illnesses (0.8\%) because they are unlikely to be related to any practice rule changes but tend to spike and drop quickly, introducing noise into the data. This left a total of N=7,425 injuries over ten seasons. 

\subsection{Outcome}
\label{sec:outcome}

The primary outcomes of interest in these analyses were conditioning-dependent and non-conditioning-dependent injury counts at the season and player-season levels. We used counts rather than rates because the new CBA reduced practices and thus decreased the total number of athlete-exposures (AEs), a common denominator unit in calculating rates of sports injuries. This reduction could increase injury rates even if it decreased the number of injuries overall. To examine injury severity we investigated the number of regular season games missed due to conditioning-dependent, non-conditioning-dependent, and all injuries as a secondary outcome. 

We defined an injury as any relevant event that appeared on the public injury reports NFL teams release before each regular season game or that placed a player on the long-term “injured reserve” list. A new injury had to meet one of two criteria: 1. It was to a different location (e.g. foot, ankle) from any previous injury that season, or 2. It was to the same location as a previous injury that season, but the player had not been on the injury report with that injury for at least two weeks, excluding weeks without a game. 

\textit{Conditioning-Dependent Injuries}: The classification of injuries as conditioning or non-conditioning was done by an orthopedics and sports medicine physician with extensive experience as an NFL team physician. Full lists of conditioning and non-conditioning injuries are given in Table A1 in the Appendix. Conditioning injuries involved soft tissues such as the Achilles tendon, calf, groin, hamstring, biceps, triceps, pectoral, quadriceps, and anterior cruciate ligament (ACL). Non-conditioning injuries included contact injuries such as fractures, high ankle sprains, and various types of trauma to the face and eye, Lisfranc joint, internal organs, neck, ribs, or toes. Some injuries – such as non-ACL knee and ankle injuries – could not be reliably categorized as conditioning-dependent or non-conditioning-dependent and were kept as a separate category or only included in the all-injury analysis. We conducted a sensitivity analysis re-classifying all non-ACL knee and ankle injuries as non-conditioning. We also conducted a sensitivity analysis using only hamstring injuries, which are some of the most likely injuries to be impacted by poor conditioning.

\subsection{Exposure}
\label{sec:exposure}

The exposure of interest was the CBA rule, which went into effect between the 2010 and 2011 seasons. As a result our exposure is perfectly defined by time, which can be split into a post-CBA (2011-2016) and a pre-CBA (2007-2010) period.

\subsection{Statistical Analysis}
\label{sec:stats}

The primary hypothesis for the NFL’s practice restrictions worsening injury risks is that with less training players are more poorly conditioned, leading to more injuries. We thus sought to evaluate the CBA’s effects within a subset of injuries most likely to be impacted by this mechanism of poor conditioning and in those that weren’t. In the absence of any bias, the CBA training restrictions could plausibly affect injury risk via three possible mechanisms:

1. Fewer practices $\xrightarrow{}$ more rest $\xrightarrow{}$ fewer injuries

2. Fewer practices $\xrightarrow{}$ poorer conditioning $\xrightarrow{}$ more injuries

3. Fewer practices $\xrightarrow{}$ fewer chances for injury $\xrightarrow{}$ fewer injuries

Each of the three proposed mechanisms is expected to differentially affect counts of conditioning and non-conditioning injuries \autoref{table1}.

\begin{table}[h!]
  \begin{center}
    \caption{Possible Effects of CBA on Conditioning and Non-Conditioning Injuries.}
    \label{table1}
    \begin{tabular}{ |p{2in}|p{2in}|p{2in}| } 
      \hline
      \textbf{CBA's Effect on Conditioning Injuries} & \textbf{CBA's Effect on Non-Conditioning Injuries} & \textbf{Mechanisms Responsible}\\
      \hline
      Fewer Injuries & Fewer Injuries & 1, 3\\
      Fewer Injuries & No Change/More Injuries & 1\\
      No Change & Fewer Injuries & 3\\
      No Change & No Change/More Injuries & None, or 1 and 2 offsetting\\
      More Injuries & Fewer Injuries & 2, 3\\
      More Injuries & No Change/More Injuries & 2\\
      \hline
    \end{tabular}
  \end{center}
\end{table}

To examine the hypothesis that CBA practice restrictions may affect risk of injury we first calculated counts of injuries and games missed by year. We then stratified these counts by whether the injuries were conditioning-dependent. We inspected these curves for evidence of immediate and delayed effects of the CBA on injury counts. A rise or fall was defined as a sustained change of at least 1 injury per 1,000 AEs, which is equivalent to a 6.7\% change in injury counts, for three or more seasons from the pre-CBA to post-CBA period.

In order to account for a secular trend in increasing injuries before the 2011 CBA we used a mixed Poisson interrupted time series model at the player-season level to separate out the effects of the CBA from broader time trends:
\\

\centerline{$ln(Y_{ij})= ln(G_{ij}) + \beta_0 + \beta_1*t_{ij} + \beta_2*CBA_{ij} + \beta_3*PostCBA_{ij} + \beta_4*Age_{ij} + b_{0i} + e_{ij}$}

where $Y_{ij}$ = count of injuries or games missed due to injury for the i’th player in the j’th season; $t_{ij}$ = Year – 2007; $CBA_{ij}$ = 1 if 2011 or later, 0 if 2010 or earlier; $PostCBA_{ij}$ = 0 for 2007-11, otherwise Year – 2011; $Age_{ij}$ = age of the i’th player in the j’th season; and $b_{0i}$ is a random intercept to account for correlated risks of injury within the same player. $ln(G_{ij})$ is an offset accounting for the number of games player i was at risk of injury in season j. When exponentiated, $\beta_2$ represents the effect of the CBA on injuries (i.e. the percent change in injury rates from the pre-CBA to post-CBA eras). When exponentiated, $\beta_1$ represents the percent change in injury rates from one year to another in the pre-CBA era. When exponentiated, $\beta_1 + \beta_3$ represents the percent change in injury rates from one year to another in the post-CBA era. To assess model fit we summed each player-season’s predicted count of injuries or games missed due to injury for each year and plotted these predicted counts against the actual injury counts for each year. 

\textit{Sensitivity analyses}: We conducted four sensitivity analyses to test the robustness of the above analysis. They involved: 1. including minor (non-game-loss) injuries; 2. including preseason injuries; 3. re-classifying knee and ankle injuries from unknown conditioning status to non-conditioning; and 4. examining only hamstring injuries, which we expected to be especially impacted by a player’s conditioning.

All analyses were performed in R version 3.3.2 and RStudio version 1.0.143 with the exception of the Poisson models, which were run in SAS v. 9.4 (SAS Institute, Cary, NC). The Emory University IRB determined this project was not human subjects research as all data is publicly available.

\section{Results}

\subsection{Overview of Time Trends}
\label{sec:results1}

\begin{figure}
  \includegraphics[width=\linewidth]{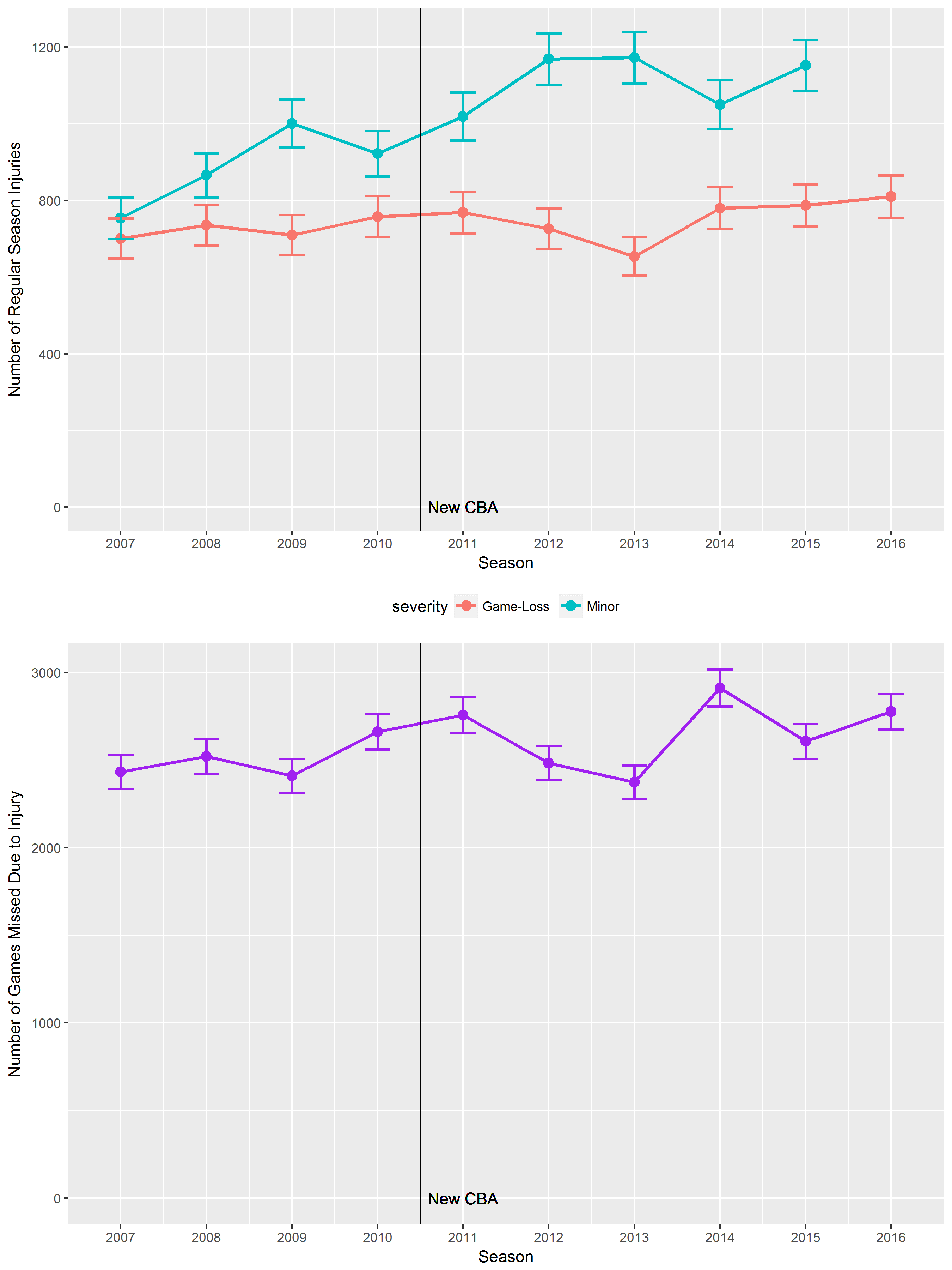}
  \caption{Number of Regular Season Non-Head, Non-Illness Injuries and Weeks Missed, 2007-2016, with 95\% CIs.}
  \label{Fig1}
\end{figure}

The number of minor regular season, non-head, non-illness injuries increased substantially from 754 in 2007 to 1,169 in 2012 (55\% rise) from 2007-2012 before stabilizing through 2016 (\autoref{Fig1}, blue line). The number of game-loss injuries exhibited a smaller increase of approximately 15\% (804 in 2016 vs. 701 in 2007). The number of regular season weeks missed (\autoref{Fig1}, purple line) – which accounts for injury severity – largely corresponds to the number of game-loss injuries (\autoref{Fig1}, red line). 

\subsection{Time Trends Stratified by Conditioning Status}
\label{sec:results2}

\begin{figure}
  \includegraphics[width=\linewidth]{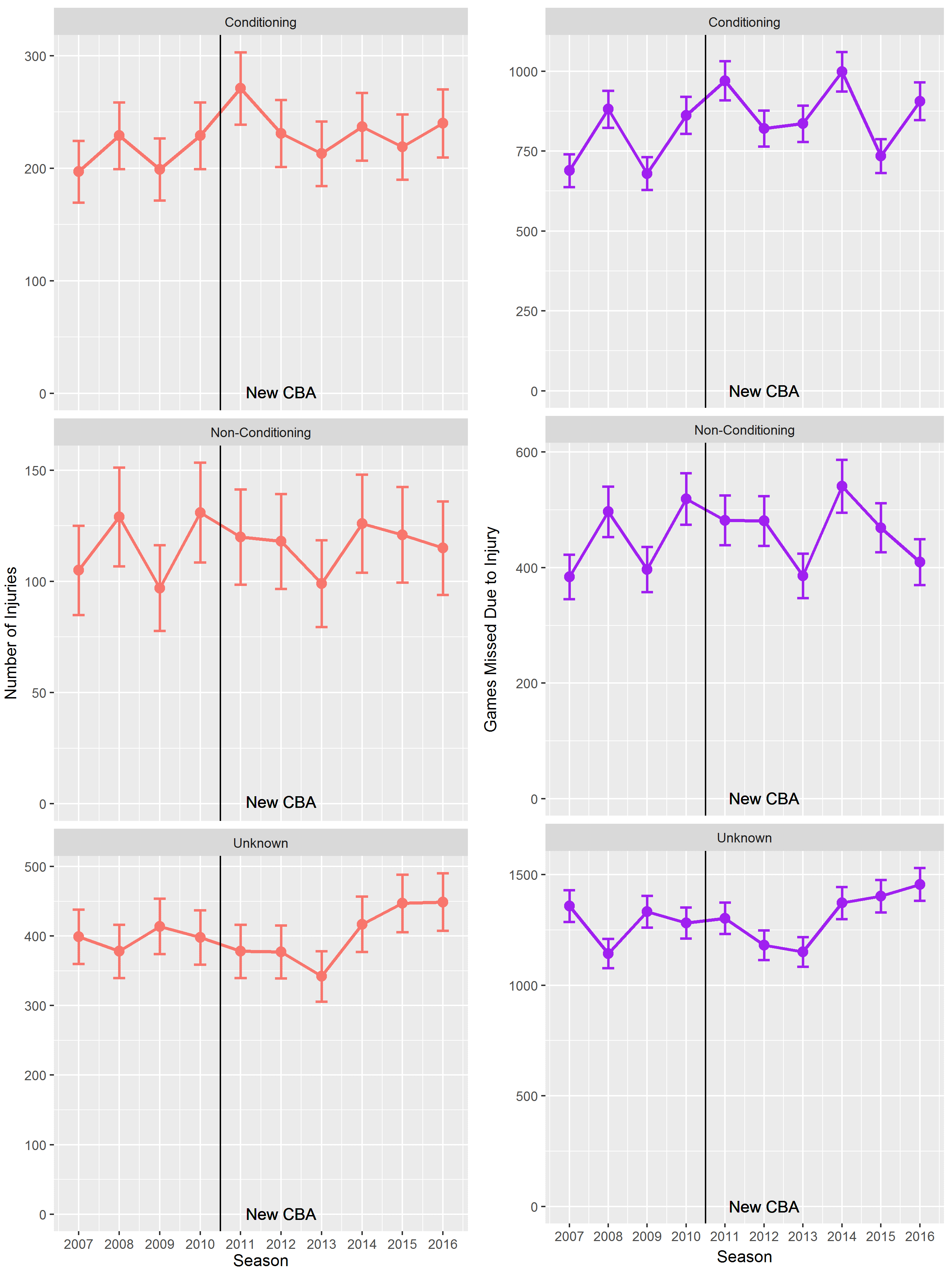}
  \caption{Number of Regular Season Game-Loss, Non-Head, Non-Illness Injuries and Games Missed, Stratified by Conditioning Status, 2007-2016, with 95\% Poisson CIs.}
  \label{Fig2}
\end{figure}

 The overall number of regular season, non-head, non-illness conditioning-dependent injuries increased from 197 in 2007 to 271 in 2011 (38\% rise) from before reaching a plateau at 220-240 injuries per season between 2012 and 2016 (\autoref{Fig2}, top left). Regular season non-conditioning injuries remained at historical levels in post-CBA (\autoref{Fig2}, middle left). Other regular season injuries remained stable before rising 19\% between 2012 and 2015 (\autoref{Fig2}, lower left). The time trends in the number of regular season weeks missed resemble those of the injury counts (\autoref{Fig2}, right). 
 
\subsection{Poisson Interrupted Time Series Models}
\label{sec:results3}

Among all regular season non-head, non-illness injuries there was little evidence of a detrimental effect of the 2011 CBA (\autoref{Fig1}, top row, red line). A Poisson model estimated that injury rates were 7\% lower in the post-CBA era than they would have been had the CBA never been implemented (95\% CI -17\% to +3\%) (\autoref{table2}, “CBA” coefficient). The model also estimated an annual 3\% increase in injury rates in both the pre- (95\% CI 0\% to +7\%) and post-CBA (95\% CI +1\% to +4\%) eras (\autoref{table2}). For total games missed, the model estimated a pre-CBA annual increase in the rate of games missed of 13\% (95\% CI +11\% to +16\%), while in the post-CBA period this decreased to a 4\% annual rise (95\% CI +3\% to +5\%) (\autoref{table2}). 

Among conditioning-dependent injuries there appeared to be an immediate one-year increase post-CBA (\autoref{Fig2}, top row); the model estimated that injury rates were 5\% higher overall in the post-CBA era than the pre-CBA era (95\% CI -13\% to +27\%) (\autoref{table2}). These injuries appeared to be increasing prior to the CBA before stabilizing in the post-CBA era (\autoref{Fig2}, top row). This result was consistent with the Poisson model, which estimated a pre-CBA annual change in injury rates of 4\% (95\% CI -2\% to +10\%) but no time trend in the post-CBA era (-1\% annual change, 95\% CI -4\% to +2\%) (\autoref{table2}). The results were similar when considering the number of games missed due to injury (\autoref{table2}).

Non-conditioning injuries remained at historical levels (\autoref{Fig2}, middle row). The model estimated that non-conditioning injury rates were 10\% lower in the post-CBA era than they would have been had the CBA never been implemented (95\% CI -31\% to +16\%). It also estimated an annual increase in the injury rate of 5\% CBA (95\% CI -4\% to +14\%) and no time trend in the post-CBA era (0\% annual change, 95\% CI -4\% to +5\%) (\autoref{table2}). The results were similar when considering the number of games missed due to injury (\autoref{table2}).

\begin{table}[h!]
  \begin{center}
    \caption{Poisson Models for Regular Season Game-Loss, Non-Head, Non-Illness Injuries and Games Missed, Stratified by Conditioning Status, 2007-2016.}
    \label{table2}
    \begin{tabular}{ |p{3in}|p{1in}|p{1in}| } 
      \hline
      \textbf{Model} & \textbf{Rate Ratio} & \textbf{95\% CI}\\
      \hline
      \multicolumn{3}{|c|}{\textbf{Number of Injuries}}\\
      \hline
      \textit{All Injuries} & & \\
      \hspace{3mm}Pre-CBA Time Trend (1-year increase) & 1.03 & 1.00, 1.07\\
      \hspace{3mm}CBA (Post-CBA vs. Pre-CBA) & 0.93 & 0.83, 1.03\\
      \hspace{3mm}Post-CBA Time Trend (1-Year increase) & 1.03 & 1.01, 1.04\\
      \hspace{3mm}Age (1-year increase) & 1.01 & 1.00, 1.01\\
      \textit{Conditioning Injuries} & & \\	 	 
      \hspace{3mm}Pre-CBA Time Trend (1-year increase) & 1.04 & 0.98, 1.10\\
      \hspace{3mm}CBA (Post-CBA vs. Pre-CBA) & 1.05 & 0.87, 1.27\\
      \hspace{3mm}Post-CBA Time Trend (1-Year increase) & 0.99 & 0.96, 1.02\\
      \hspace{3mm}Age (1-year increase) & 1.02 & 1.00, 1.03\\
      \textit{Non-Conditioning Injuries} & & \\	 	 
      \hspace{3mm}Pre-CBA Time Trend (1-year increase) & 1.05 & 0.96, 1.14\\
      \hspace{3mm}CBA (Post-CBA vs. Pre-CBA) & 0.90 & 0.69, 1.16\\
      \hspace{3mm}Post-CBA Time Trend (1-Year increase) & 1.00 & 0.96, 1.05\\
      \hspace{3mm}Age (1-year increase) & 1.02 & 1.00, 1.03\\
      
      \hline
      \multicolumn{3}{|c|}{\textbf{Number of Games Missed Due to Injury}}\\
      \hline
      \textit{All Injuries} & & \\
      \hspace{3mm}Pre-CBA Time Trend (1-year increase) & 1.13 & 1.11, 1.16\\
      \hspace{3mm}CBA (Post-CBA vs. Pre-CBA) & 0.90 & 0.85, 0.96\\
      \hspace{3mm}Post-CBA Time Trend (1-Year increase) & 1.04 & 1.03, 1.05\\
      \hspace{3mm}Age (1-year increase) & 1.07 & 1.06, 1.08\\
      \textit{Conditioning Injuries} & & \\	 	 
      \hspace{3mm}Pre-CBA Time Trend (1-year increase) & 1.13 & 1.09, 1.18\\
      \hspace{3mm}CBA (Post-CBA vs. Pre-CBA) & 0.97 & 0.88, 1.08\\
      \hspace{3mm}Post-CBA Time Trend (1-Year increase) & 1.04 & 1.02, 1.07\\
      \hspace{3mm}Age (1-year increase) & 1.10 & 1.08, 1.11\\
      \textit{Non-Conditioning Injuries} & & \\	 	 
      \hspace{3mm}Pre-CBA Time Trend (1-year increase) & 1.19 & 1.13, 1.25\\
      \hspace{3mm}CBA (Post-CBA vs. Pre-CBA) & 0.90 & 0.78, 1.03\\
      \hspace{3mm}Post-CBA Time Trend (1-Year increase) & 0.98 & 0.95, 1.01\\
      \hspace{3mm}Age (1-year increase) & 1.06 & 1.04, 1.08\\
      \hline
    \end{tabular}
  \end{center}
\end{table}

\subsection{Sensitivity Analyses}
\label{sec:results4}

\begin{figure}[!ht]
  \includegraphics[width=\linewidth]{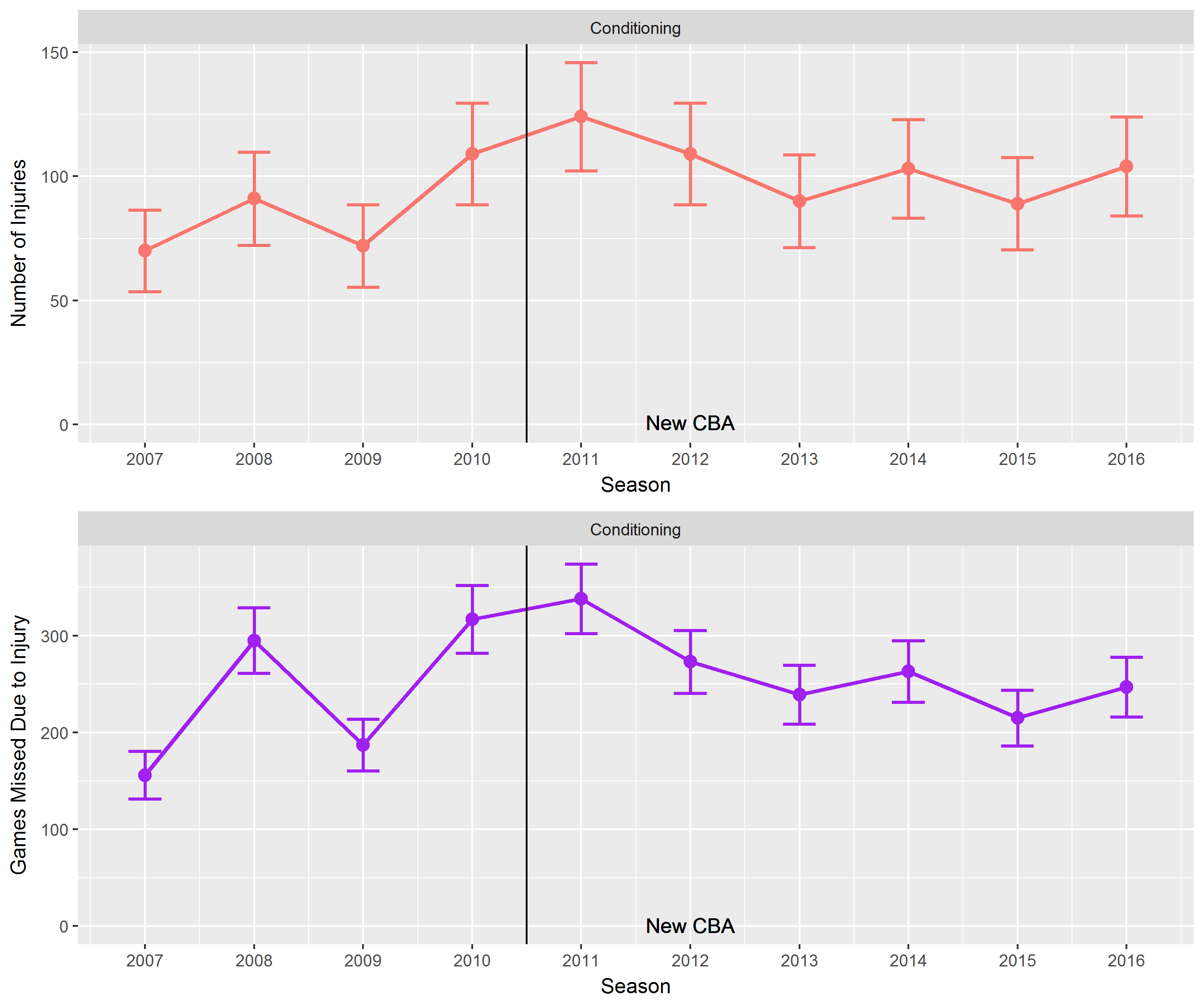}
  \caption{Number of Regular Season Game-Loss Hamstring Injuries, 2007-2016, with 95\% Confidence Interval.}
  \label{Fig3}
\end{figure}

We conducted several sensitivity analyses. Including minor injuries in did not substantially alter the model’s coefficients, nor did including preseason injuries or re-classifying knee and ankle injuries from unknown to non-conditioning. When examining hamstring injuries specifically, we observed an increase of these injuries over time pre-CBA followed by a reversing of that trend in the post-CBA era (\autoref{Fig3}). The Poisson models also indicated stronger beneficial effects of the CBA. The models estimated a pre-CBA annual increase in hamstring injury rates of 12\% (95\% CI +2\% to +23\%) but a 4\% annual decrease in the post-CBA era (95\% CI -8\% to +1\%) (\autoref{table3}). The results were similar when considering the number of games missed due to injury (\autoref{table3}).

\begin{table}[!ht]
  \begin{center}
    \caption{Poisson Models for Regular Season Game-Loss Hamstring Injuries and Games Missed, 2007-2016.}
    \label{table3}
    \begin{tabular}{ |p{3in}|p{1in}|p{1in}| } 
      \hline
      \textbf{Model} & \textbf{Rate Ratio} & \textbf{95\% CI}\\
      \hline
      \textit{Hamstring Injury Counts} & & \\
      \hspace{3mm}Pre-CBA Time Trend (1-year increase) & 1.12 & 1.02, 1.23\\
      \hspace{3mm}CBA (Post-CBA vs. Pre-CBA) & 0.98 & 0.74, 1.31\\
      \hspace{3mm}Post-CBA Time Trend (1-Year increase) & 0.96 & 0.92, 1.01\\
      \hspace{3mm}Age (1-year increase) & 0.97 & 0.95, 0.99\\
      \textit{Hamstring Games Missed} & & \\	 	 
      \hspace{3mm}Pre-CBA Time Trend (1-year increase) & 1.29 & 1.21, 1.38\\
      \hspace{3mm}CBA (Post-CBA vs. Pre-CBA) & 0.73 & 0.61, 0.88\\
      \hspace{3mm}Post-CBA Time Trend (1-Year increase) & 0.99 & 0.95, 1.02\\
      \hspace{3mm}Age (1-year increase) & 1.01 & 0.99, 1.03\\
      \hline
    \end{tabular}
  \end{center}
\end{table}

\section{Discussion}

Overall regular season game-loss, non-head, non-illness injuries did not rise or fall from the pre- to post-CBA practice restriction period. Our conclusions are primarily based on the descriptive data in \autoref{Fig1} and \autoref{Fig2} and the definitions of rise and fall presented in the Methods, but they are consistent with the interrupted time series models presented in \autoref{table2}, which show no evidence for an increase in conditioning-dependent injuries or injuries overall in the post-CBA era. In the context of the three injury-affecting mechanisms we outlined in the Methods, the results suggest we have seen either none of these mechanisms operating or nearly-offsetting effects from additional rest/fewer chances for injury and poorer conditioning. 

The limitations of the Poisson interrupted time series models underlying \autoref{table2} merit additional discussion. As with any interrupted time series analysis our conclusions are reliant on assumed counterfactuals for the post-CBA period. If we assume that injury counts would have continued rising unabated the CBA’s practice restrictions may appear beneficial. If we instead assume that injury rates would have plateaued even in the absence of the CBA then the effects could better be described as non-detrimental. In models that included no pre- or post-CBA time trends, the CBA appeared detrimental (results not presented). The choice of counterfactual assumption exerts a strong effect on the interpretations of our models.

With that said, the models are a useful supplement to visual graph inspection and descriptive analyses. Among all regular season non-head, non-illness injuries visual inspection and comparison to the rise/fall criteria outlined in the Methods suggested no detrimental effect of the practice restrictions (\autoref{Fig1}, top row, red line). Consistent with our interpretation, the model estimated injury rates were 7\% lower in the post-CBA era than they would have been absent the CBA and no change in time trends pre- and post-CBA (\autoref{table2}). Visual analysis suggested a similar lack of detrimental effects for conditioning-dependent injuries (\autoref{Fig2}, top row). The Poisson model did not identify a substantial difference in injury rates or time trends in the pre- vs. post-CBA eras (\autoref{table2}). Visual inspection of non-conditioning injuries suggested they remained largely at historical levels (\autoref{Fig2}, middle row). This interpretation is consistent with the results of the Poisson models, though the large year-to-year variance in these injuries (\autoref{Fig2}) suggests the model’s coefficients may not be reliable. 

In 2014 there was a substantial jump in injuries of unknown conditioning status that lasted through the 2016 season (\autoref{Fig3}, bottom row). The increase was across a range of injury locations – knee, ankle, foot, back, and shoulder, primarily. This is unlikely to be an effect of the CBA’s practice restrictions due to its delayed timing and sudden onset. However, we were unable to identify a single event or set of events between the 2013 and 2014 seasons to account for this change. 

To our knowledge this is the first study to investigate the injury effects of the 2011 CBA’s practice restrictions over a substantial time period. One research group identified a sudden increase in Achilles injuries immediately after the 2011 NFL lockout ended and training camp began (12 Achilles tendon ruptures the first 29 days after the lockout versus 6 and 10 total Achilles tendon ruptures the last two full seasons, respectively) \cite{Myer11}. Our findings are consistent with theirs, however: we classified Achilles injuries as conditioning-dependent, and such injuries did exhibit a temporary bump in 2011 before returning closer to historical levels (\autoref{Fig2}).

This study has several strengths. We analyzed 10 full years of data – 4 years pre- and 6-years post-intervention – which allows us to place the effects of the practice restrictions in context of broader injury trends and mitigates the risks of drawing conclusions from natural season-to-season variations. We also attempted to account for time trends that might distort a simple pre-/post-CBA comparison using interrupted time series models.

There are also several important limitations in this study. First, there were several other changes designed to enhance player safety and reduce injuries concomitant with the 2011 CBA and throughout the study period. Changes concomitant with the CBA included moving the kickoff up to encourage more touchbacks and the expansion of the “defenseless players” list to include, among others, receivers who have not re-established themselves as runners. It is difficult to disentangle the effects of all these various changes in our data, but it is most likely these effects would have biased injuries in later years downward from what they would have been absent these changes. This would in turn have made the CBA look more beneficial than it truly was. 

Second, we interpreted our models assuming as a counterfactual that the pre-CBA rise in injuries would have continued unabated during our study period absent the CBA. If instead injuries would have plateaued absent the CBA our estimated effects would be biased in favor of the CBA.

Third, our data often only gives us the body part injured rather than the specific injury, which along with the unknown impact of conditioning on many injuries inserts the possibility for misclassification of conditioning/non-conditioning injuries. Additional information such as whether the injury was contact or non-contact would help with making these designations in future studies. The direction of this bias can be inferred by comparing the results for all conditioning injuries (\autoref{Fig2}, top row) to those for only hamstring injuries (\autoref{Fig3}), which are known to be affected by over- and undertraining \cite{Carey17}. The trend in hamstring injuries switches from an increase pre-CBA to a decrease post-CBA (\autoref{Fig3} and \autoref{table3}); if this represents the true effect of practice restrictions on conditioning-dependent injuries, then our full conditioning-dependent results (which do not show decreasing injury counts post-CBA) may suffer from misclassification that biases it against a beneficial effect of the CBA.

\section{Conclusions}

Among regular season game-loss, non-head, non-illness injuries, descriptive analyses and interrupted time series models did not indicate the CBA’s practice restrictions had a net harmful effect on injury burden in the NFL. It does not appear that the practice restrictions pushed the NFL, on average, from a state of optimal training to one of undertraining, but whether there was previous overtraining – or whether there is still undertraining – remains unclear. However, other concurrent injury-related rule and regulation changes are potential confounding factors that limit specific causal inferences about the practice restrictions, and further studies are warranted.

\bibliography{Aim2_CBA.bib} 
\bibliographystyle{vancouver}

\newpage

\appendix
\renewcommand\thefigure{\thesection.\arabic{figure}}
\renewcommand\thetable{\thesection.\arabic{table}}
\section{Appendix}
  \setcounter{figure}{0}
  \setcounter{table}{0}

\subsection{Conditioning Injury Definitions}
\label{sec:appendix1}

\autoref{tableA1} lists 51 injury types from our injury database and how they were classified with respect to the impact of conditioning on the incidence of those injuries. As described below, we also conducted a sensitivity analysis where ankle sprains, general ankle injuries, knee sprains, and general knee injuries were classified as not conditioning-dependent rather than unknown. These categories were chosen because they were the largest components of the unknown group, which was substantially larger than either the conditioning or non-conditioning groups in the main analysis.

\begin{longtable}{ |p{2in}|p{2in}|p{2in}| }
    \caption{Condensed Injury Types and Conditioning Status.}
    \label{tableA1}\\
      \hline
      \textbf{Condensed Injury Type} & \textbf{Conditioning-Dependent?} & \textbf{Alternative Conditioning Definition}\\
      \hline
      \endfirsthead
      \endhead
      Abdomen & No &\\	
      Achilles & Yes &\\	
      Ankle - High Ankle Sprain & Unknown &\\
      Ankle - Other & Unknown & No\\
      Ankle - Sprain & Unknown	& No\\
      Arm - Broken & No &\\ 	
      Arm - Other & No &\\ 
      Back & Unknown &\\ 	
      Biceps & Yes &\\	
      Buttocks & Yes &\\ 	
      Calf & Yes &\\ 	
      Chest & Yes &\\ 	
      Elbow & No &\\ 	
      Eye & No &\\	
      Face & No &\\ 	
      Finger & No &\\ 	
      Foot - Broken & No &\\ 	
      Foot - Lisfranc & No &\\ 	
      Foot - Other & Unknown &\\ 	
      Groin & Yes &\\ 	
      Hamstring & Yes &\\ 	
      Hand & No &\\ 	
      Head - Concussion & EXCLUDED &\\ 	
      Head - Other & EXCLUDED &\\ 	
      Heart & No &\\ 	
      Hip & Unknown &\\ 	
      Illness & EXCLUDED &\\ 	
      Knee - ACL & Yes &\\ 
      Knee - Other & Unknown & No\\
      Knee - Sprain & Unknown & No \\
      Knee - Tear, non-ACL & Unknown &\\ 	
      Leg - Broken & No &\\ 	
      Leg - Other & No &\\ 	
      Lung & No &\\ 	
      Neck & No &\\ 	
      Non-Injury & No &\\ 	
      Other & No &\\ 	
      Pectoral & Yes &\\ 	
      Pelvis & Unknown &\\ 	
      Quadriceps & Yes &\\ 	
      Ribs & No &\\ 	
      Shoulder - Other & Unknown &\\ 	
      Shoulder - Tear & Unknown &\\ 	
      Spine & Unknown &\\ 	
      Thigh & Unknown &\\ 	
      Thumb & No &\\ 	
      Toe - Other & No &\\ 	
      Toe - Turf Toe & No &\\ 	
      Triceps & Yes &\\ 	
      Unknown & Unknown &\\ 	
      Wrist & No &\\ 
      \hline
\end{longtable}

\subsection{Stratified Time Trends}
\label{sec:appendix2}

\begin{figure}
  \includegraphics[width=\linewidth]{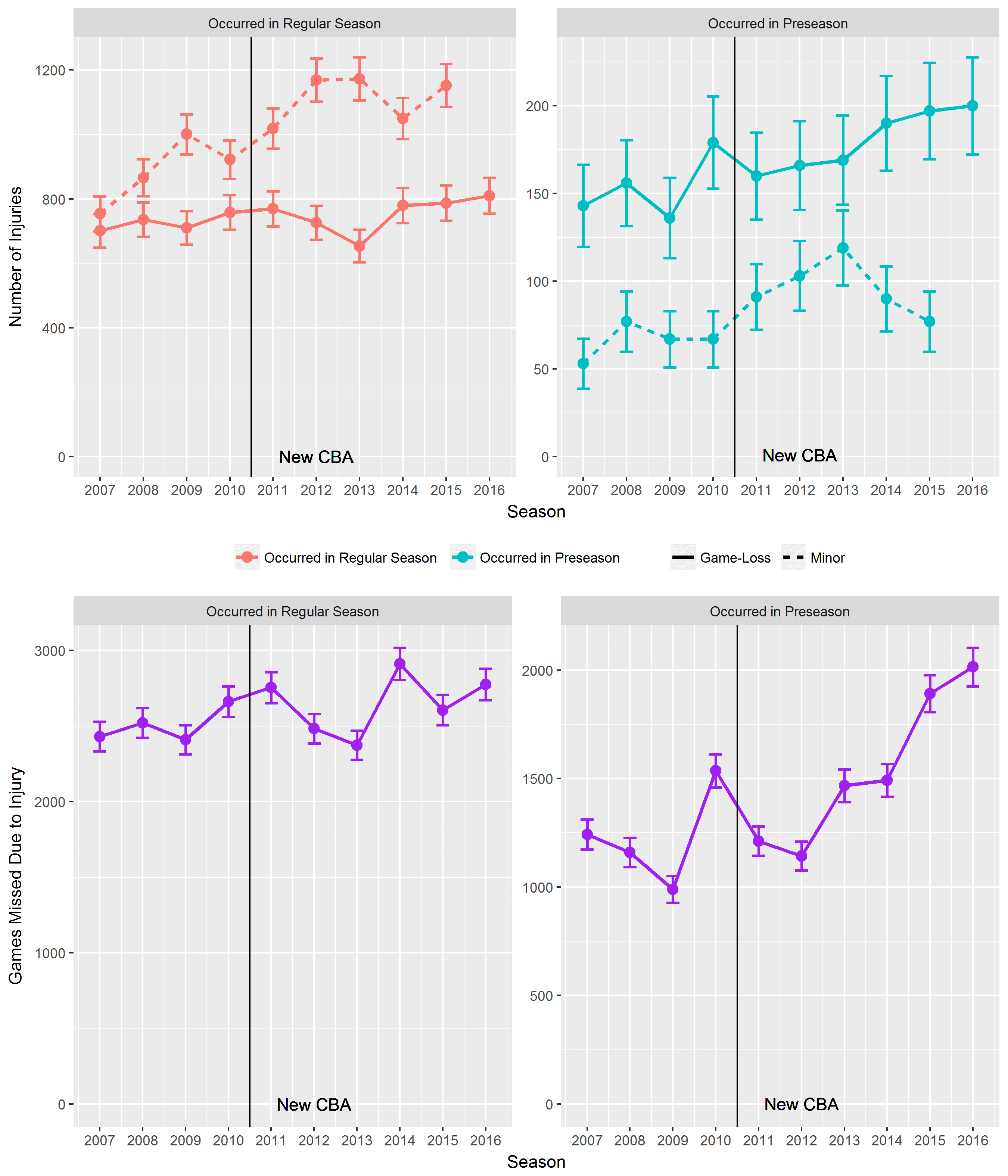}
  \caption{Number of Regular Season Game-Loss Hamstring Injuries, 2007-2016, with 95\% Confidence Interval.}
  \label{FigA1}
\end{figure}

Among regular season injuries, there has been a large increase in minor injuries while game-loss injuries have remained relatively flat over our study period (\autoref{FigA1}, upper left). Among preseason injuries, both minor and game-loss injuries exhibited steady increases over our study period (\autoref{FigA1}, upper right).

When looking at games missed due to injury, those due to injuries in the regular season exhibited a modest but inconsistent increase over our study period (\autoref{FigA1}, lower left). Games lost due to preseason injuries, however, exhibited a rapid increase from 2012-2016 (\autoref{FigA1}, lower right). The spike from 2014 to 2015-16 was almost entirely driven by an increase in “Undisclosed” injuries that landed players on injured reserve, the majority of which occurred to career backups who were likely at the end of their careers. These may not reflect a true increase so much as more thorough reporting and capturing of these sorts of injuries.

\subsection{Sensitivity Analysis – Include Preseason Injuries}
\label{sec:appendix3}

As seen in \autoref{FigA1}, preseason game-loss injuries and the games missed due to them increased steadily from 2007-2015 while game-loss injuries occurring in the regular season remained relatively flat. When we include preseason injuries in our analysis our injury count results do not differ substantially from the main analysis (\autoref{tableA2}, \autoref{table2}).

While it would be ideal to include preseason injuries in our analysis as this may be where the CBA’s practice restrictions had their strongest effect, it is difficult to define a player population that is exchangeable with the pre-CBA period due to another concomitant rule instituted by the CBA. Specifically, the maximum training camp roster increased from 80 players to 90; the cut-down periods during training camp have also been progressively pushed back in recent years. All this means more players are collecting more exposures to injury in the preseason, driving counts up. We attempted to account for this by limiting our target population to players with at least one career NFL game, but there are still a substantial number of veterans with game experience who make these expanded training camp rosters but would not have made a training camp in the pre-CBA period \cite{Duffy16}. We attempted but could not find a satisfactory way to define a consistent population of preseason players that is exchangeable between the pre- and post-CBA periods. Thus we have limited our main analysis to the regular season, which has maintained a consistent 53-man roster throughout the entire study period. “Practice squad” rosters – which contains players who usually do not play except as injury replacements – did expand from 8 to 10 under the new CBA, as well, but the bias from this is likely minor and would be in the direction of a detrimental effect of the new CBA. 

\begin{table}[h!]
  \begin{center}
    \caption{Poisson Models for Regular Season Game-Loss, Non-Head, Non-Illness Injuries and Games Missed, Stratified by Conditioning Status, 2007-2016.}
    \label{tableA2}
    \begin{tabular}{ |p{3in}|p{1in}|p{1in}| } 
      \hline
      \textbf{Model} & \textbf{Rate Ratio} & \textbf{95\% CI}\\
      \hline
      \multicolumn{3}{|c|}{\textbf{Number of Injuries}}\\
      \hline
      \textit{All Injuries} & & \\
      \hspace{3mm}Pre-CBA Time Trend (1-year increase) & 1.04 & 1.01, 1.07\\
      \hspace{3mm}CBA (Post-CBA vs. Pre-CBA) & 0.94 & 0.85, 1.03\\
      \hspace{3mm}Post-CBA Time Trend (1-Year increase) & 1.02 & 1.00, 1.03\\
      \hspace{3mm}Age (1-year increase) & 1.00 & 0.99, 1.01\\
      \textit{Conditioning Injuries} & & \\	 	 
      \hspace{3mm}Pre-CBA Time Trend (1-year increase) & 1.05 & 1.00, 1.12\\
      \hspace{3mm}CBA (Post-CBA vs. Pre-CBA) & 1.01 & 0.85, 1.20\\
      \hspace{3mm}Post-CBA Time Trend (1-Year increase) & 0.99 & 0.96, 1.02\\
      \hspace{3mm}Age (1-year increase) & 1.01 & 1.00, 1.02\\
      \textit{Non-Conditioning Injuries} & & \\	 	 
      \hspace{3mm}Pre-CBA Time Trend (1-year increase) & 1.05 & 0.98, 1.14\\
      \hspace{3mm}CBA (Post-CBA vs. Pre-CBA) & 0.89 & 0.70, 1.12\\
      \hspace{3mm}Post-CBA Time Trend (1-Year increase) & 1.00 & 0.96, 1.05\\
      \hspace{3mm}Age (1-year increase) & 1.00 & 0.99, 1.02\\

      \hline
      \multicolumn{3}{|c|}{\textbf{Number of Games Missed Due to Injury}}\\
      \hline
      \textit{All Injuries} & & \\
      \hspace{3mm}Pre-CBA Time Trend (1-year increase) & 1.21 & 1.19, 1.24\\
      \hspace{3mm}CBA (Post-CBA vs. Pre-CBA) & 0.85 & 0.80, 0.89\\
      \hspace{3mm}Post-CBA Time Trend (1-Year increase) & 0.98 & 0.97, 0.99\\
      \hspace{3mm}Age (1-year increase) & 1.10 & 1.09, 1.11\\
      \textit{Conditioning Injuries} & & \\	 	 
      \hspace{3mm}Pre-CBA Time Trend (1-year increase) & 1.28 & 1.23, 1.32\\
      \hspace{3mm}CBA (Post-CBA vs. Pre-CBA) & 0.85 & 0.78, 0.93\\
      \hspace{3mm}Post-CBA Time Trend (1-Year increase) & 1.00 & 0.98, 1.02\\
      \hspace{3mm}Age (1-year increase) & 1.11 & 1.09, 1.12\\
      \textit{Non-Conditioning Injuries} & & \\	 	 
      \hspace{3mm}Pre-CBA Time Trend (1-year increase) & 1.26 & 1.21, 1.32\\
      \hspace{3mm}CBA (Post-CBA vs. Pre-CBA) & 0.77 & 0.68, 0.87\\
      \hspace{3mm}Post-CBA Time Trend (1-Year increase) & 0.96 & 0.94, 0.99\\
      \hspace{3mm}Age (1-year increase) & 1.05 & 1.03, 1.07\\
      \hline
      
    \end{tabular}
  \end{center}
\end{table}

\end{document}